# High-pressure phase transitions of zinc difluoride up to 55 GPa

Dominik Kurzydłowski[1*], Anna Oleksiak[1], Sharad Babu Pillai,[2] Prafulla K. Jha[2]

[1] *Faculty of Mathematics and Natural Sciences, Cardinal Stefan Wyszyński University, Warsaw 01-038, Poland;*

[2] *Department of Physics, Faculty of Science, The Maharaja Sayajirao University of Baroda, Vadodara 390002, India*

***Correspondence:*** *d.kurzydlowski@uksw.edu.pl*

Studying the effect of high pressure (exceeding 10 kbar) on the structure of solids allows to gain deeper insight in the mechanism governing crystal structure stability. Here we report a study on the high-pressure behaviour of zinc difluoride ($ZnF_2$) – an archetypical ionic compound which at ambient pressure adopts the rutile ($TiO_2$) structure. Previous investigations, limited to a pressure of 15 GPa, revealed that this compound undergoes two pressure-induced phase transitions: $TiO_2$ → $CaCl_2$ at 4.5 GPa, and $CaCl_2$ → HP-$PdF_2$ at 10 GPa. Within this joint experimental-theoretical study we extend the room-temperature phase diagram of $ZnF_2$ up to 55 GPa. By means of Raman spectroscopy measurements we identify two new phase transitions: HP-$PdF_2$ → HP1-$AgF_2$ at 30 GPa and HP1-$AgF_2$ → $PbCl_2$ at 44 GPa. These results are confirmed by Density Functional Theory calculations which indicate that in the HP1-$AgF_2$ polymorph the coordination sphere of $Zn^{2+}$ undergoes drastic changes upon compression. Our results point to important differences in the high-pressure behaviour of $ZnF_2$ and $MgF_2$, despite the fact that both compounds contain cations of similar size. We also argue that the HP1-$AgF_2$ structure, previously observed only for $AgF_2$, might be observed at large compression in other $AB_2$ compounds.

## Introduction

Rationalizing the crystal structures of solids is an ongoing endeavour in material sciences. Concepts such as the Goldshimdt tolerance factor,[1] Pauling rules,[2] or the bond-valence index[3] are important tools for understanding and predicting crystal geometry. Metal difluorides ($MF_2$) are an excellent group of compounds for testing these and other concepts.[4,5] One reason for this is the large variation of the $M^{2+}$ cationic radius which increases from 0.45 Å for $Be^{2+}$ to 1.35 Å for $Ba^{2+}$ (Shannon's effective ionic radius for six-fold coordination, $R_{M^{2+}}$).[6] Another important feature of $MF_2$ systems lies in the range of electronic configurations adopted by $M^{2+}$ cations: for group 2 metals they exhibit a closed-shell, for group 12 and 14 a filled sub-shell (*d* for group 12, *s* and *d* for group 14), while transition metal cations have an open *d* shell.

In recent years particular attention was drawn to phase transitions of $MF_2$ compounds induced by pressures exceeding 1 GPa (=10 kbar). Numerous experimental and theoretical investigations were conducted in order to establish the phase sequence for these compounds.[7–25] The motivation for these studies was the fact that the behaviour of difluorides at these conditions can serve as an analogue of the high-pressure transformations of dioxides, in particular $SiO_2$.[7,8] Moreover high-pressure studies on $MF_2$ systems revealed a surprisingly wide range of structures adopted by these compounds. At ambient conditions difluorides adopt structures belonging either to the rutile ($TiO_2$) or fluorite ($CaF_2$) family, with the exception of $BeF_2$ and $SnF_2$.[26,27] In contrast, at high pressure the potential energy surface of $MF_2$ systems is characterized by a number of local minima of similar enthalpy, often separated by large energy barriers.[18]

For transition metal difluorides it was found that the high-pressure phase transition sequence differs considerably between ions of similar size, but with different $d$ electron count, as witnessed by comparing $MnF_2$ ($d^5$ system)[8,19] and $CoF_2$ ($d^7$ electron count).[20] Moreover for $d^9$ cations ($Cu^{2+}$, $Ag^{2+}$) the high-pressure polymorphism is strongly influenced by structural distortion arising from an uneven occupation of the $e_g$ orbitals (often referred to as the Jahn-Teller effect)[28] as well as strong spin-spin magnetic interactions.[24] In this context the phase transitions of zinc difluoride are of interest, as the $Zn^{2+}$ cation can be viewed as a non-magnetic ($3d^{10}$ electronic configuration) counterpart of $Mn^{2+}$, $Co^{2+}$, and $Cu^{2+}$. Given the similar size of $Zn^{2+}$ and $Mg^{2+}$ ($R_{Zn^{2+}}$ =0.74 Å, $R_{Mg^{2+}}$ =0.72 Å) one could expect that the phase transition of $ZnF_2$ would be similar to those exhibited by $MgF_2$.[18] However experimental data for the former compound is only available to a moderate pressure of 15 GPa,[9,10,25] while theoretical predictions were restricted to pressures not exceeding 12 GPa, and performed taking into account only a limited number of possible structure types.[11,12]

Here we extend the phase diagram of $ZnF_2$ to 55 GPa by performing Raman scattering measurements with the use of the diamond anvil cell (DAC). The obtained data confirms the previously reported phase transition from a $CaCl_2$-type structure to the HP-$PdF_2$ polytype at 10.4 GPa.[10] Two additional phase transitions are observed at 30 and 44 GPa. We assign the first of these as a transformation from the HP-$PdF_2$ polymorph to a non-centrosymmetric structure (space group $Pca2_1$, $Z = 4$) analogous to the high-pressure structure of $AgF_2$ (HP1-$AgF_2$ structure type). The second transition at 44 GPa is between HP1-$AgF_2$ and the cotunnite ($PbCl_2$) structure. Our results are confirmed by Density Functional Theory (DFT) calculations and evolutionary algorithm structure searches. DFT modelling indicates that upon compression of the HP1-$AgF_2$ polymorph the coordination sphere of $Zn^{2+}$ smoothly evolves from octahedral to tricapped trigonal prismatic. Our results indicate also important differences in the high-pressure behaviour of $MgF_2$ and $ZnF_2$.

## Experimental and computational details

**Raman spectroscopy:** Raman spectra were acquired at room temperature with the use of the Alpha300M+ confocal microscope (Witec Gmbh). We used a 532 nm laser line delivered to the microscope through a single-mode optical fiber. Laser power at the sample did not exceed 30 mW. The Raman signal was collected through a 20× long working distance objective, and passed through a multi-mode optical fiber (50 μm core diameter) to a lens-based spectrometer (Witec UHTS 300, f/4 aperture, focal length 300 mm) coupled with a back-illuminated Andor iDUS 401 detector thermoelectrically cooled to –60°C. The spectra were collected with the use of a 1800 mm grating resulting in a 1.2 cm$^{-1}$ spectral resolution. Typical acquisition times ranged from 1 to 4 s with 30 to 60 accumulations. The spectra were post-processed (background subtraction and cosmic-ray removal) with the Project FIVE software (Witec Gmbh). The position of Raman bands was established with the Fityk 1.3.1 software by fitting the observed bands with Pseudo-Voigt profiles.[29]

**High-pressure experiments:** two high-pressure runs were conducted with the use of a SymmDAC diamond anvil cell supplied by Almax easyLab bvba. The DAC was equipped with low-fluorescence Ia diamonds with a 400 μm culet and a stainless-steel gasket pre-indented to a thickness of 35 μm. The gasket hole with a radius of 120 μm was laser-drilled. The hole was filled by zinc difluoride powder (anhydrous, 99% purity, supplied by Thermo Fisher Gmbh) after which the cell was closed. The starting pressure of the experiments was 4.5 GPa. The pressure was determined from the shift of the R1 ruby fluorescence line.[30] During the experiments the DAC was systematically heated to 100°C in order to minimize deviatoric stress in the sample.

**DFT calculations:** Periodic DFT calculations of the geometry and enthalpy of various polymorphs of $ZnF_2$ utilized the SCAN meta-GGA functional.[31] This functional was found to be superior to both Local Density Approximation (LDA) and Generalized-Gradient Approximation (GGA) functionals in predicting the phase transition pressures for a wide range of compounds.[32,33] SCAN was also successfully used to model the properties of the high-pressure phases of silver difluoride, $AgF_2$.[24] We found it to reproduce very well the geometry of the previously reported polymorphs of $ZnF_2$ (see Table S1 in the Supplementary Information, SI). The projector-augmented-wave (PAW) method was used in the calculations,[34] as implemented in the VASP 5.4 code.[35,36] The cut-off energy of the plane waves was set to 800 eV with a self-consistent-field convergence criterion of $10^{-7}$ eV. Valence electrons (Zn: $3d^{10}$, $4s^2$; F: $2s^2$, $2p^5$) were treated explicitly, while standard VASP pseudopotentials were used for the description of core electrons. The k-point mesh spacing was set to $2\pi \times 0.03$ Å$^{-1}$. All structures were optimized until the forces acting on the atoms were smaller than 1 meV/Å.

Calculations were performed for 14 structure types listed in Table 1. Additionally we performed evolutionary algorithm searches for lowest-enthalpy structures of $ZnF_2$. For this we used the XtalOpt software (version r12)[37] coupled with periodic DFT calculations utilizing the PBE functional.[38] These searches were conducted at 10/30/80 GPa for $Z = 4$, and 35/50 GPa for $Z = 8$.

Calculations of Γ-point vibration frequencies were conducted in VASP 5.4 utilizing the SCAN functional. The finite-displacement method was used with a 0.025 Å displacement, and a tighter SCF convergence criterion ($10^{-8}$ eV). In case of the HP1-$AgF_2$, $SrI_2$, and $PbCl_2$ structure types we additionally calculated the intensity of Raman-active modes using density-functional perturbation theory (DFPT),[39] as implemented in Quantum Espresso code.[40] The exchange-correlation functional was handled by the LDA approximation of Perdew and Wang.[41] Expansion of wave function and charge density in the plane wave basis set was truncated with energy cut-offs of 110, and 170 Ry for HP1-$AgF_2$/$PbCl_2$ and $SrI_2$ structures, respectively. Brillouin zone was sampled on dense k-mesh ($2\pi \times 0.03$ Å$^{-1}$) given by the Monkhorst-Pack scheme. Raman activity of phonon modes were determined implementing the second order response method.[42] We did not apply any scaling of the theoretical vibration frequencies when comparing them with experimental values.

For calculations of the electronic band gap ($E_g$) of the most stable structure of $ZnF_2$ we employed the Heyd-Scuseria-Ernzerhof (HSE06) functional,[43] which is a hybrid functional mixing the GGA functional of Perdew et al.,[38] with 25% of the Hartree-Fock exchange energy. Calculations were performed for the SCAN-relaxed structures. We found that the use of the HSE06 functional for the SCAN geometries results in small forces on atoms (<0.05 eV/Å) and moderate residual stress (<5 GPa).

Visualization of all structures was performed with the VESTA software package.[44] For symmetry recognition we used the FINDSYM program.[45] Group theory analysis of the vibrational modes was performed with the use of the Bilbao Crystallographic Server.[46]

## Results and discussion

We start by introducing the structure types relevant to this study which we group into four families (see Table 1): rutile, fluorite, zirconia and cotunnite. The structures in the rutile family exhibit octahedral coordination by F$^-$ with the coordination number (CN) of the cation equal to 6. At ambient conditions the majority of known metal difluorides adopt the rutile structure ($TiO_2$, *I4/mmm*, $Z = 2$). Both $CrF_2$ and $CuF_2$ adopt a distorted variant of this structure ($CuF_2$, *P2$_1$/c*, $Z = 2$) with the first coordination sphere of $Cr^{2+}$/$Cu^{2+}$ in the form of an elongated octahedron (4 + 2 coordination).[47] The distortion leading to the $CuF_2$-type structure is a result of a negative force constant for the $B_{2g}$ and $B_{3g}$ vibration modes of the $TiO_2$ structure.[48,49] Compression of fluorides exhibiting the $TiO_2$ structure often results in a second-order ferroelastic phase transition to an orthorhombic structure (CaCl$_2$, *Pnnm*, *Z* = 2), as seen for MgF$_2$ at 9.1 GPa.[18] This phase, which differs from the TiO$_2$ polytype by rotation of the MF$_6$ octahedra along axes parallel to the *c* lattice vector, retains the coordination environment of M$^{2+}$. The last member of the rutile family is the α-PbO$_2$ structure (*Pbcn*, *Z* = 4) which also exhibits octahedral coordination of M$^{2+}$.

**Table 1** Structure types adopted by MF$_2$ systems together with their space group, number of formula units per cell (*Z*), and the coordination number of the M$^{2+}$ cation (*CN*).

| Structure family | Structure type | Space group | Z | CN | Comments |
|---|---|---|---|---|---|
| Rutile | TiO$_2$ | *P4$_2$/mnm* | 2 | 6 | rutile |
| | CuF$_2$ | *P2$_1$/c* | 2 | 4 + 2 | distorted TiO$_2$ |
| | CaCl$_2$ | *Pnnm* | 2 | 6 | distorted TiO$_2$ |
| | α-PbO$_2$ | *Pbcn* | 4 | 6 | |
| Fluorite | CaF$_2$ | *Fm-3m* | 4 | 8 | fluorite |
| | HP-PdF$_2$ | *Pa-3* | 4 | 6 | |
| | AgF$_2$ | *Pbca* | 4 | 4 + 2 | distorted HP-PdF$_2$ |
| | HP1-AgF$_2$ | *Pca2$_1$* | 4 | * | distorted AgF$_2$ |
| Zirconia | ZrO$_2$ | *P2$_1$/c* | 4 | 7 | baddeleyite |
| | SrI$_2$ | *Pbca* | 8 | 7 | brookite, orthorhombic-I |
| | HS-ZrO$_2$ | *Pbcm* | 4 | 8 | |
| | HP2-AgF$_2$ | *Pbcn* | 8 | 4 + 4 | distorted HS-ZrO$_2$ |
| Cotunnite | PbCl$_2$ | *Pnma* | 4 | 9 | cotunnite |
| | Ni$_2$In | *P6$_3$/mmc* | 2 | 11 | |

\* coordination number dependent on pressure (*vide infra*).

The second family of structures is derived from the fluorite polytype (CaF$_2$, *Fm-3m*, *Z* = 4). This structure, adopted at ambient conditions by several MF$_2$ systems (*e.g.* CaF$_2$, SrF$_2$, and BaF$_2$), is characterized by an eightfold (cubic) coordination of M$^{2+}$. In the HP-PdF$_2$ structure (*Pa-3*, *Z* = 4) the first coordination sphere of the metal is a distorted octahedron and there are two secondary contacts which complete the cubic coordination. This polytype is observed as the high-pressure form of many MF$_2$ systems that at ambient conditions adopt the TiO$_2$ structure (*e.g.* MgF$_2$ has the following phase sequence: TiO$_2$ $\xrightarrow{9.1\ GPa}$ CaCl$_2$ $\xrightarrow{14\ GPa}$ HP-PdF$_2$).[18] The HP-PdF$_2$ polytype has the same space group and Wyckoff position sequence as pyrite (FeS$_2$), but does not exhibit any F-F bonding analogous to S-S bonding present in FeS$_2$.

The remaining two polytypes in the fluorite family are connected with the polymorphs of silver difluoride. At ambient conditions this compound adopts a lower-symmetry variant of the HP-PdF$_2$ structure (AgF$_2$, *Pbca*, *Z* = 4) with the first coordination sphere of Ag$^{2+}$ in a form of an elongated octahedron.[50] This structure transforms at around 9 GPa to a non-centrosymmetric polymorph (HP1-AgF$_2$, *Pca2$_1$*, *Z* = 4).[22] The transformation is driven by a phonon instability of the AgF$_2$ structure.

The third family of structures is derived from the polymorphs of zirconia (ZrO$_2$). At ambient pressure and temperature the most stable polymorph of this compound has monoclinic symmetry (ZrO$_2$, *P2$_1$/c*, *Z* = 4). In this structure the metal cation is 7-fold coordinate in the form of a monocapped trigonal prism (capping of one of the rectangular faces). There is one longer secondary contact along the other rectangular face of the trigonal prism. Upon compression ZrO$_2$ transforms into a SrI$_2$-type structure (*Pbca*, *Z* = 8) at 10 GPa.[51] This structure retains the monocapped trigonal prism coordination, but with two additional secondary contacts. The phase transition between the ZrO$_2$ and SrI$_2$ structure is first order. In a theoretical study a metastable (*i.e.* thermodynamically less stable than the SrI$_2$-type structure) high-pressure polymorph of ZrO$_2$ was proposed which is characterized by *Pbcm* symmetry and *Z* equal to 4.[52] This polymorph results from the pressure-induced symmetrisation of the ZrO$_2$-type structure and hence we term it as HS-ZrO$_2$. The metal sites in HS-ZrO$_2$ are 8-fold coordinated in a form of a bicapped trigonal prism (capping at two rectangular faces); there is one additional contact at a longer distance. The last member of the zirconia family is the structure-type derived from the second high-pressure polymorph of AgF$_2$ (HP2-AgF$_2$, *Pbcn*, *Z* = 8) adopted by this compound above 15 GPa.[21,22] This structure can be viewed as a distorted variant of HS-ZrO$_2$ with the first coordination sphere split into four shorter and four longer contacts.

All transition metal difluorides that adopt the CaF$_2$ structure at ambient conditions transform to the cotunnite structure (PbCl$_2$, *Pnma*, *Z* = 4) at relatively low pressures (< 10 GPa).[7,13–16] This structure is characterized be an nine-fold coordination of M$^{2+}$ in the form of a tricapped trigonal prism with the three anions capping the rectangular faces. At higher pressures the PbCl$_2$ polytype transforms to an Ni$_2$In structure in which two additional anions enter the coordination sphere of M$^{2+}$ leading to a 11-fold coordination (pentacapped trigonal prism), the highest observed for metal difluorides.[7]

**High-pressure Raman scattering**

As nearly all of the first-row transition metal difluorides, zinc difluoride (ZnF$_2$) adopts the TiO$_2$ structure at ambient conditions.[53] Raman measurements on ZnF$_2$ compressed to 6.5 GPa in a diamond anvil cell (DAC) indicated that this compound undergoes a transformation to the CaCl$_2$-type polymorph at 4.5 GPa.[9] This transition was confirmed by x-ray diffraction measurements which also indicated a second transition to a HP-PdF$_2$ structure (space group *Pa-3*) commencing at 10 GPa.[10]

Upon decompression from 15.3 GPa the HP-PdF$_2$ polymorph was stable down to 4 GPa. Below this pressure the sample transformed to a mixture of two phases: TiO$_2$ and α-PbO$_2$-type.[10]

The Raman spectrum of ZnF$_2$ in the range 4.5 to 10 GPa (Figure 1) is in accordance with that previously measured for the CaCl$_2$-type polymorph.[9] From the six Raman-active modes (2A$_g$ + 2B$_{1g}$ + B$_{2g}$ + B$_{3g}$) two modes of A$_g$ symmetry are observed together with a band resulting from the overlap of one of the B$_{1g}$ modes with a B$_{2g}$ and a B$_{3g}$ mode. The assignment of the modes is based on the comparison between the observed band positions and the frequencies of the Raman-active modes calculated with the use of the SCAN functional (see Figure S 1). The one remaining B$_{1g}$ mode, predicted to appear at frequencies above 550 cm$^{-1}$ is not observed, most probably due to its low intensity.

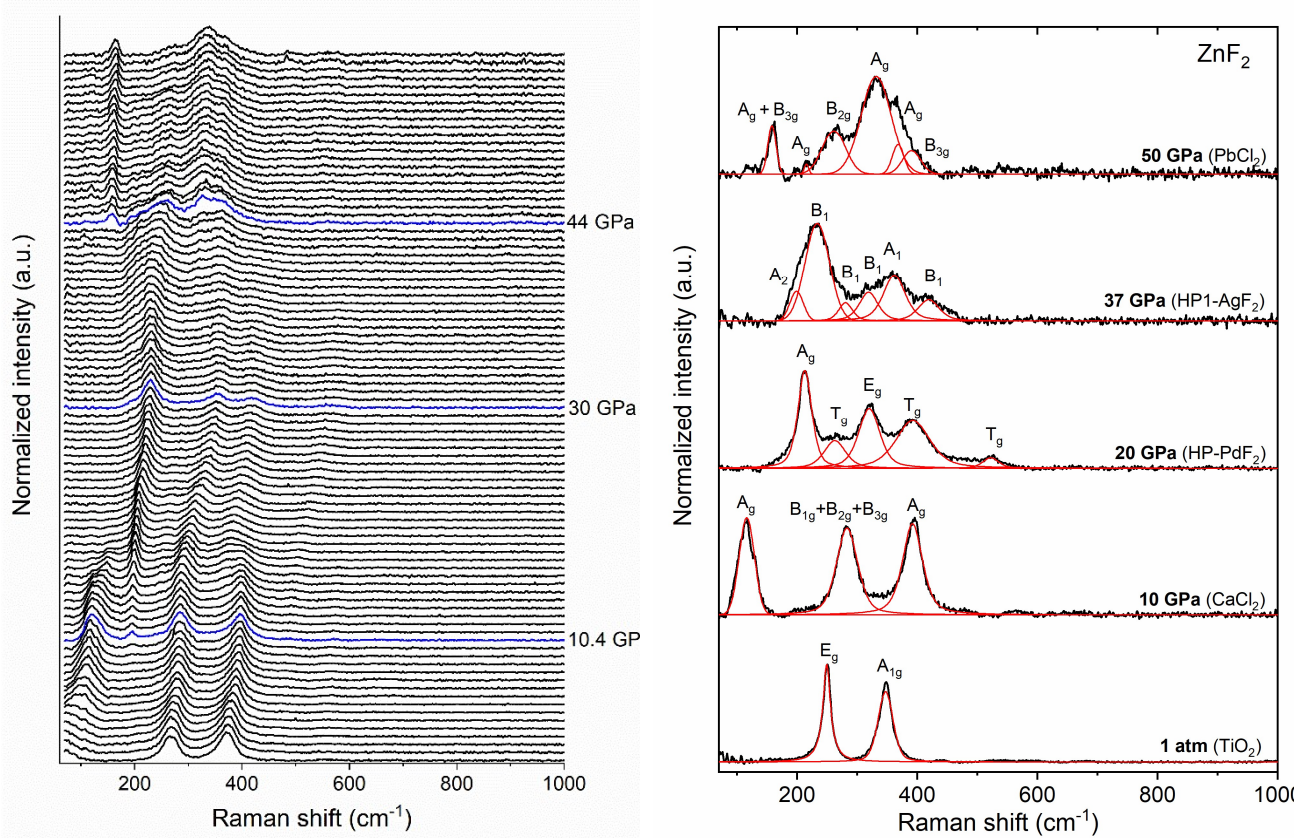

**Figure 1 Left:** Evolution of the Raman spectrum of powdered ZnF$_2$ from 4.5 to 55 GPa (the spectra are offset for clarity). Spectra marked in blue are taken at pressures at which phase transitions are observed (10.4, 30 and 40 GPa). **Right:** Raman spectrum of ZnF$_2$ at selected pressure together with the deconvolution into component functions. Labels denote the symmetry of each mode.

Upon compression of the CaCl$_2$-type phase all of the observed bands shift to higher wavenumbers (Figure 2), with a particularly large pressure dependence observed for the least energetic A$_g$ transition (Table S 2). At 10.4 GPa a new band appears around 195 cm$^{-1}$ and upon compression grows in intensity at the expense of the two A$_g$ modes (Figure 1). At the same pressure the band originating from the $B_{1g}/B_{2g}/B_{3g}$ modes changes its pressure dependence. These changes, together with the appearance of three additional bands (250, 380, and 500 cm$^{-1}$ at 13.5 GPa) mark the gradual transition from the CaCl$_2$-type polymorph to the HP-PdF$_2$ structure. This transition is completed at 15.9 GPa as signalled by the complete disappearance of the bands originating from the CaCl$_2$ polymorph. The pressure at which the CaCl$_2$ – HP-PdF$_2$ transition is observed as well as the coexistence of both phases up to 16 GPa is in accordance with previous x-ray diffraction measurements.[10] We observe all five of the Raman-active modes of the HP-PdF$_2$ structure ($A_g + E_g + 3T_g$). Thanks to an excellent agreement between measured band position and calculated frequencies of vibrational modes (see Figure S 2) we were able to assign the symmetry of all of these modes.

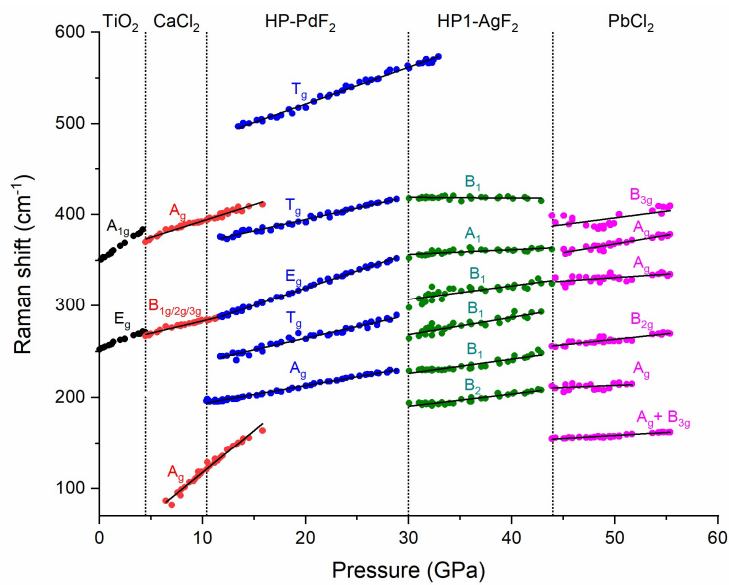

**Figure 2** Pressure dependence of the frequencies of Raman bands of ZnF$_2$ measured upon compression. Points corresponding to the TiO$_2$/CaCl$_2$/HP-PdF$_2$/HP1-AgF$_2$/PbCl$_2$ phases are marked with black/red/blue/green/magenta. Black lines are linear fits (see Table S 2). Data for the TiO$_2$ polymorph is taken from ref.[9].

Compression of the HP-PdF$_2$ polymorph to 30 GPa leads to a gradual stiffening of all of the observed modes with the highest-frequency $T_g$ mode having the largest pressure coefficient *dv/dp* (Table S 2). At 30 GPa two shoulders develop on both sides of the $A_g$ band and the intensity of the highest-frequency $T_g$ mode decreases upon compression (Figure 1 and Figure S 3). This is accompanied by a decrease in the *dv/dp* coefficient for the bands located at 355 and 420 cm$^{-1}$ at 30 GPa, as seen in Figure 2. These gradual transformations point towards a phase transition into a structure resembling the HP-PdF$_2$ polymorph, but exhibiting lower symmetry. Based on the comparison of calculated and experimental vibration frequencies (Figure S 4) we assign the phase appearing at 30 GPa to the HP1-AgF$_2$ structure. Of the 33 Raman-active modes present for this non-centrosymmetric structure ($8A_1 + 9A_2 + 8B_1 + 8B_2$) we observe 6 ($A_1 + A_2 + 3B_1$).

The last phase transition is observed at 44 GPa. Its most characteristic feature is the redistribution of intensity of the band observed in the 200 – 400 cm$^{-1}$ region, and the appearance of a band at 150 cm$^{-1}$ whose position varies only slightly with pressure (*dv/dp* equal to 0.65 cm$^{-1}$/GPa). Comparison of the calculated and experimental frequencies of Raman active modes (Figure S 5) indicated that the phase observed above 44 GPa is the PbCl$_2$-type (cotunnite) polymorph of ZnF$_2$. With the aid of the SCAN calculations we are able to assign the observed Raman bands. From the 18 Raman-active modes (6A$_g$ + 3B$_{1g}$ + 6B$_{2g}$ + 3B$_{3g}$) we observe six (4A$_g$ + B$_{2g}$ + 2B$_{3g}$) with A$_g$ and B$_{3g}$ low-frequency modes coalescing into one band at 150 cm$^{-1}$. This analysis is corroborated by the good accordance between the experimental spectrum and that simulated for the PbCl$_2$ polymorph (Figure S 5).

**DFT calculations**

In order to confirm the nature of the experimentally observed phase transitions we calculated the geometry and enthalpy of various phases of ZnF$_2$ at pressures up to 100 GPa. For these calculations we assumed the structure types listed Table 1. Additional evolutionary algorithm searches were performed in order to identify other possible high-pressure polymorphs of ZnF$_2$ (see Materials and Methods). However these searches did not yield any new structures which would be competitive in terms of enthalpy with those listed in Table 1.

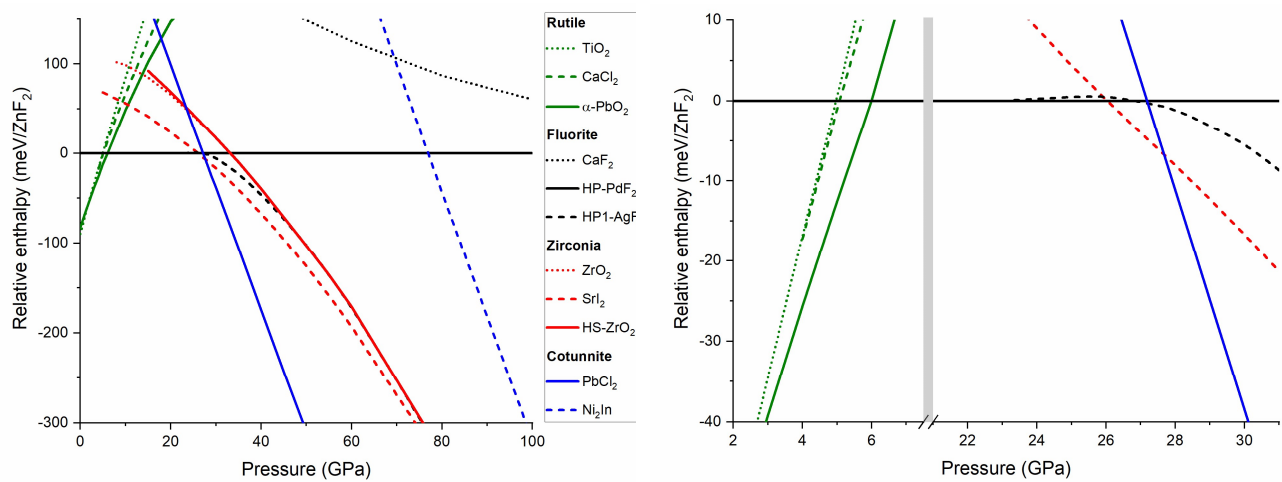

**Figure 3 Left:** Pressure dependence of the relative enthalpy (referenced to that of the HP-PdF$_2$ structure) of ZnF$_2$ polymorphs up to 300 GPa. **Right**: Zoom into the 2 – 7.5 and 21 – 31 GPa region.

Upon geometry optimization the CuF$_2$, AgF$_2$, and HP2-AgF$_2$ structure types symmetrized to the TiO$_2$, HP-PdF$_2$, and HS-ZrO$_2$ structures, respectively. This is not surprising given that the three former polymorphs are distorted variants of the latter, with the distortions arising from the uneven occupation of the *e$_g$* orbitals which is not present for the closed-shell Zn$^{2+}$ cation. Interestingly the non-centrosymmetric HP1-AgF$_2$ structure did retain its symmetry upon geometry optimization, although the coordination around the M$^{2+}$ cation changed from 4 + 2 to 6 at 25 GPa.

In accordance with experiment at ambient pressure (effectively 0 GPa) TiO$_2$ is the most stable structure. Our calculations indicate that this structure should transform into α-PbO$_2$ at 1.8 GPa. However the α-PbO$_2$ phase is not observed experimentally. This situation resembles that found in other AB$_2$ compounds (*e.g.* MgF$_2$ and SnO$_2$)[18,54] for which the α-PbO$_2$ structure is not observed during compression despite being more stable in calculations than the experimentally observed CaCl$_2$ structure. This discrepancy was rationalized by taking into account the large energetic barrier associated with the TiO$_2$ to α-PbO$_2$ transition.[18,54] In case of ZnF$_2$ calculations give the TiO$_2$ – CaCl$_2$ transition at 4 GPa (Figure **3**) in very good agreement with the experimental value of 4.5 GPa.[9] The transition from CaCl$_2$ to HP-PdF$_2$ is predicted at 5.1 GPa in good agreement with the experimental value (10.4 GPa).

Comparison of the relative enthalpies of ZnF$_2$ polymorphs indicates that the HP-PdF$_2$ polymorph remains the most stable structure of ZnF$_2$ up to a pressure of 26 GPa. Above this pressure the SrI$_2$ structure type is predicted to be the ground state, but only in a narrow pressure range up to 27.7 GPa. Above this pressure the PbCl$_2$ (cotunnite) structure has a lower enthalpy and remains the most stable polymorph up to 100 GPa.

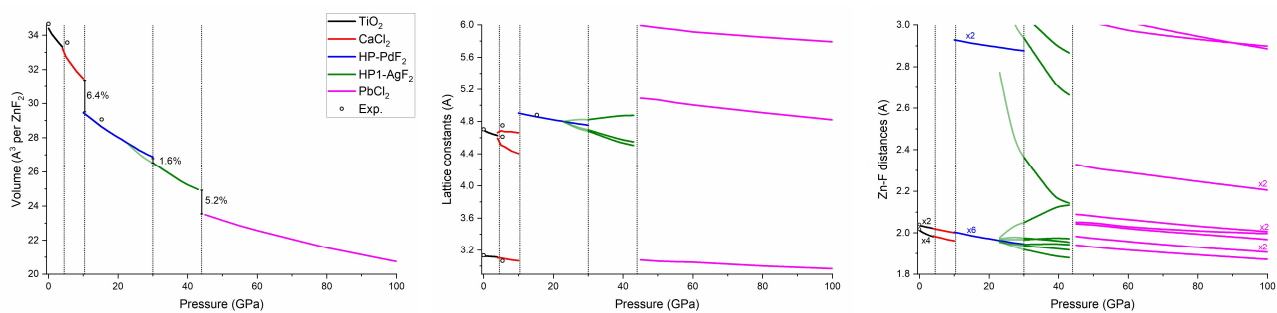

**Figure 4** Calculated pressure dependence of the volume (left), lattice constants (middle), and Zn-F distances (right) of the experimentally observed polymorphs of ZnF$_2$. Dotted vertical lines indicate experimental transition pressures (4.5, 10.4, 30, and 44 GPa). Open circles indicate experimental values.[10,53]

In experiment we observe that at 30 GPa that HP-PdF$_2$ transforms to the HP1-AgF$_2$. The latter structure is indeed more stable than HP-PdF$_2$ above 26 GPa but in the same pressure regime it is less stable than SrI$_2$ and PbCl$_2$ (Figure 3). Comparing the evolution of the cell vectors and Zn-F distances (Figure 4) one can clearly see that orthorhombic HP1-AgF$_2$ is a distorted variant of the cubic HP-PdF$_2$. Moreover calculations show that HP1-AgF$_2$ can be obtained from HP-PdF$_2$ by following an imaginary Γ-point mode of T$_u$ symmetry which develops in the cubic structure at 32 GPa. Therefore at this pressure the HP1-AgF$_2$ to HP-PdF$_2$ transition should be characterized by a very small energetic barrier. In contrast the transition from HP-PdF$_2$ to SrI$_2$ or PbCl$_2$ is most probably characterized by a large barrier as it requires a change in the geometry around the M$^{2+}$ ion from 6-fold octahedral to 7-fold monocapped trigonal prismatic (SrI$_2$) or 9-fold tricapped trigonal prismatic (PbCl$_2$). The existence of a low-energy distortion pathway between HP-PdF$_2$ and HP1-AgF$_2$ explains why this transition is observed in the room-temperature experiment. We also note that in case of MgF$_2$ the SrI$_2$ phase is also not observed experimentally,[18] although it is predicted to be thermodynamically stable between 40 and 44 GPa.[17]

In experiment the metastable HP1-AgF$_2$ phase is observed up to 44 GPa. At this point it transforms to the PbCl$_2$ structure. It may be speculated that with the use of high temperatures (T > 1000 K) this transition could be observed at lower pressures. High temperatures could also enable the direct transformation of HP-PdF$_2$ to PbCl$_2$, as witnessed in the case of MgF$_2$.[18] Such experiments would however require laser heating of the sample which is unavailable to us in the current experiment experimental set-up. Our calculations indicate that PbCl$_2$ should remain the most stable structure up to 350 GPa at which it should transform to the anti-Ni$_2$In structure which features 11-coordinated Zn$^{2+}$. The large value of the pressure required for the PbCl$_2$ to anit-Ni$_2$In transition is consistent with the small radius of Zn$^{2+}$.[7]

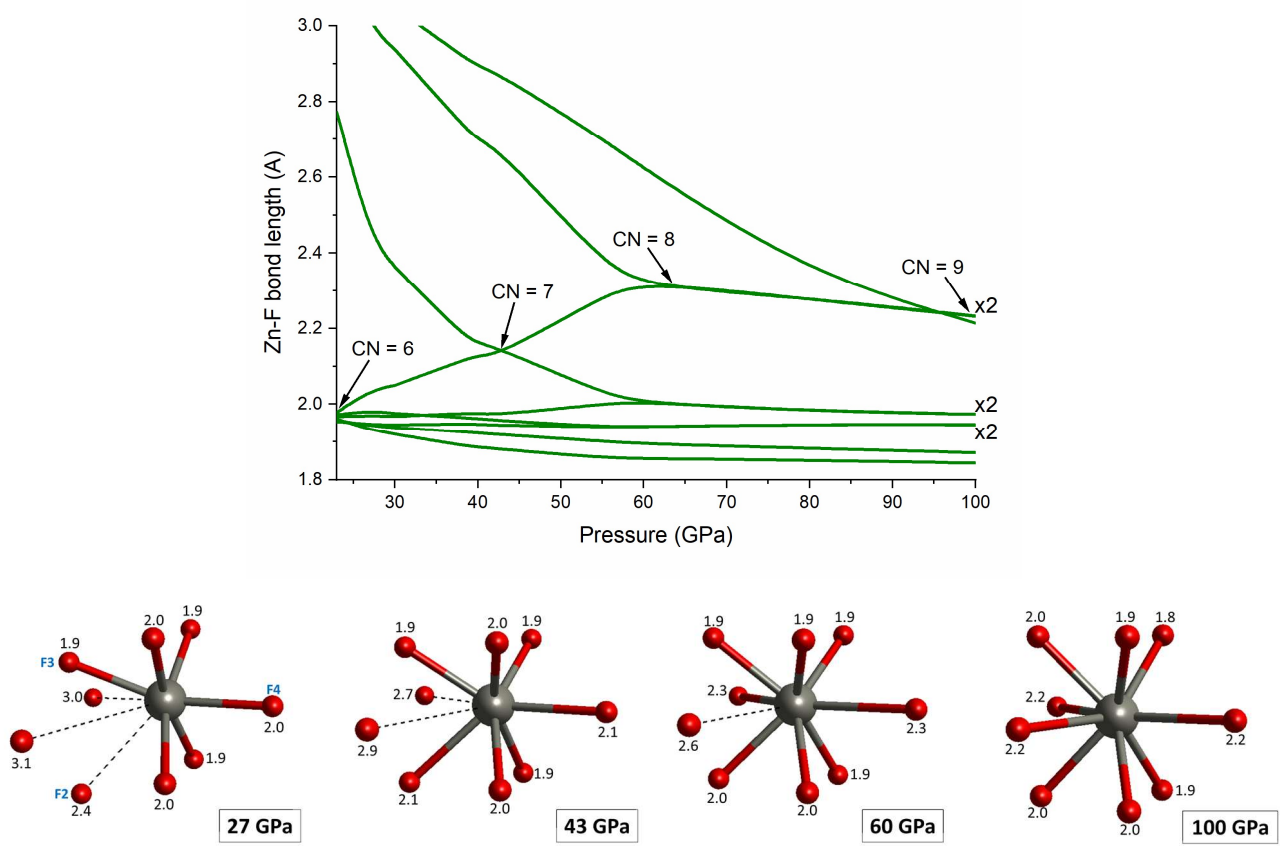

**Figure 5** Calculated pressure dependence of the Zn-F distances in the HP1-AgF$_2$ polymorph of ZnF$_2$ together with the evolution of the Zn$^{2+}$ coordination sphere in that structure. Distances are given in Å.

Calculation reveal substantial pressure-induced changes in the coordination of $Zn^{2+}$ in the HP1-AgF$_2$ structure. At 23 GPa the coordination sphere has the form of a distorted octahedron with six Zn-F contact with a distance of around 2 Å. As can be seen in Figure 5 upon compression the F3-Zn-F4 angles strongly deviates from 180°, and the Zn-F4 distances lengthens (nearly 8 % from 23 to 43 GPa). At the same time the secondary Zn-F2 contact shortens considerably (by 23 % from 23 to 43 GPa) and becomes equal in length to Zn-F4 at 43 GPa. These changes mark a gradual transition of the coordination sphere from octahedral (CN = 6) to monocapped trigonal prismatic coordination (CN = 7) which is analogous to that found in the SrI$_2$ structure (Figure 5). Further compression leads to an increase of the coordination number to 8 at 60 GPa with the coordination sphere in the form of a bicapped trigonal prism. At this pressure HP1-AgF$_2$ in fact becomes isostructural to HS-ZrO$_2$ which exhibits the same CN. Further compression leads to one more F$^-$ anion entering the first coordination sphere of $Zn^{2+}$ leading to a tricapped trigonal prismatic coordination (CN = 9) resembling that found for the PbCl$_2$ structure. The pressure-induced changes in the Zn-F distances in the HP1-AgF$_2$ polymorph (especially the lengthening of the Zn-F4 contact by 18 % from 23 to 60 GPa) points to a surprisingly large plasticity of the $Zn^{2+}$ coordination sphere resembling that found at ambient condition only for compounds containing Jahn-Teller active cations, such as $Cu^{2+}$ and $Ag^{2+}$.[55,56]

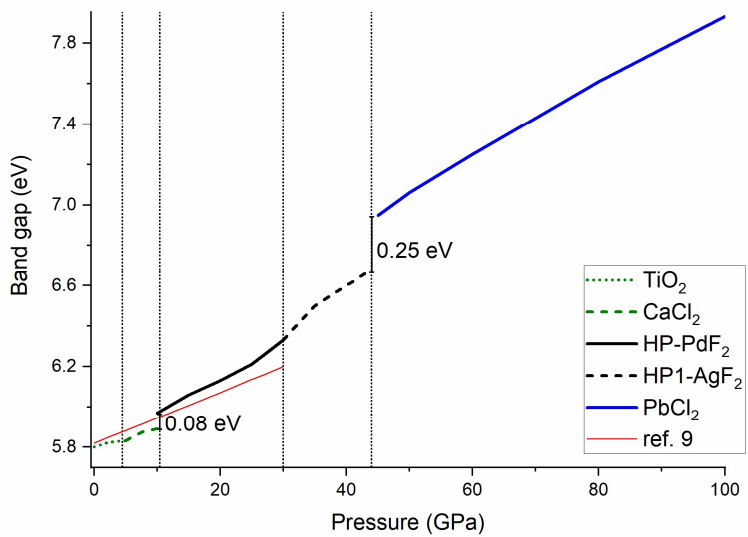

**Figure 6** The calculated electronic band gap of the experimentally observed phases of ZnF$_2$. Red line indicates values obtained for TiO$_2$/CaCl$_2$/HP-PdF$_2$ phases from GGA calculations of ref. [11] (these values are shifted by 2.2 eV for clarity). Dotted vertical lines indicate experimental transition pressures (4.5, 10.4, 30, and 44 GPa).

Calculations of the band gap of ZnF$_2$ in the TiO$_2$ structure give the value of 5.8 eV at ambient pressure. This number is smaller than the experimental value (7 – 8 eV)[57] but larger than those obtained in GGA-based calculations (3.60 – 3.65 eV).[11,58,59] For all stable ZnF$_2$ polymorphs compression leads to an increase in the band gap, but the pressure coefficient ($dE_g/dp$) differs between phases. For rutile structures ($TiO_2$, $CaCl_2$) it is equal to about 9 meV/GPa, while it is larger for the fluorite structures (21 meV/GPa) and $PbCl_2$ (18 meV/GPa). These values are larger than reported for $CoF_2$ ($dE_g/dp < 7$ meV/GPa),[20] but smaller than for $MgF_2$ ($dE_g/dp \approx 40$ meV/GPa).[17] These differences most likely stem from the different electronic configuration between $Zn^{2+}$ ($3d^{10}$), $Co^{2+}$ ($3d^7$) and $Mg^{2+}$ ($2s^22p^6$).

Comparison of the calculated phase transition sequence for $ZnF_2$ and $MgF_2$ (ref. [17]) show that for the latter compound transition pressures are shifted to higher values. This trend, which is also found when comparing experimental data,[18] can be explained by the smaller radius of $Mg^{2+}$ ($R_{Mg^{2+}}$ =0.72 Å vs. $R_{Zn^{2+}}$ =0.74 Å). A more detailed comparison of the enthalpies of the high-pressure polymorphs reveals that for $MgF_2$ the HP-$PdF_2$ and $SrI_2$ polymorphs are more stable with respect to HP1-$AgF_2$ and $PbCl_2$ than for $ZnF_2$ (Figure **7**). This difference may also be a manifestation of the different electronic structure of the cation – a notion that might be explored in future studies. The fact that for the $MgF_2$ HP1-$AgF_2$ is less competitive in term of enthalpy explains why this polymorph is not observed experimentally for this compound. We also note that when moving from $ZnF_2$ to $CuF_2$, which contains the Jahn-Teller active $Cu^{2+}$ cation ($3d^9$), the HP1-$AgF_2$ becomes even more stabilized, and is predicted to be the ground state structure of $CuF_2$ in a wide pressure range (30 – 72 GPa).[23]

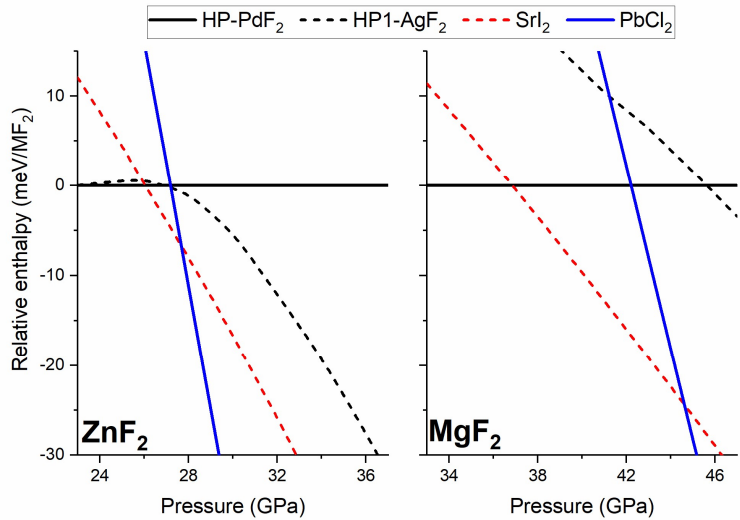

**Figure 7** Comparison of the relative enthalpies of the HP-$PdF_2$, HP1-$AgF_2$, $SrI_2$ and $PbCl_2$-type structure for $ZnF_2$ and $MgF_2$ calculated with the SCAN functional. Both graphs show a pressure interval of 14 GPa and a enthalpy window of 45 meV per $MF_2$.

## Conclusions

In summary high-pressure Raman scattering experiments conducted for $ZnF_2$ revealed two new phase transition: to the HP1-$AgF_2$ structure (space group $Pca2_1$) at 30 GPa, and to the $PbCl_2$ (cotunnite) polymorph at 44 GPa. The phase transition sequence observed for $ZnF_2$ is confirmed by DFT calculations which also unveil that the HP1-$AgF_2$ polymorph is metastable and is formed upon compression from the HP-$PdF_2$ phase through a phonon instability. The coordination sphere of $Zn^{2+}$ on the non-centrosymmetric HP1-$AgF_2$ structure changes substantially upon compression smoothly evolving from a cation coordination number of 6 to 9. Comparison of the calculated band gaps and relative phase stability between $MgF_2$ and $ZnF_2$ unearths important differences between these systems despite the nearly identical radius of $Mg^{2+}$ and $Zn^{2+}$.

The observation of the HP1-$AgF_2$ might have important implications for the study of the phase transition of other $AB_2$ compounds. We note that the cell vectors of this structure are close to being tetragonal (Figure 4) and therefore HP1-$AgF_2$ could be considered as a candidate phase for the tetragonal distorted fluorite phase of $CoF_2$.[20] HP1-$AgF_2$ exhibits a very similar x-ray diffraction pattern as $SrI_2$ (see Figure S 6 in SI) and therefore could be taken into account in the analysis of phase transition of compounds for which the $SrI_2$ phase is observed, such as $MnF_2$ or $PbO_2$.[19,60,61] Finally we note that despite the similarity of diffraction patterns the Raman spectrum of HP1-$AgF_2$ and $SrI_2$ are quite different (Figure S 7) which highlight the usefulness of Raman scattering as a tool for distinguishing different polymorphs of ionic compounds.

**Acknowledgments**: This research was carried out with the support of the Interdisciplinary Centre for Mathematical and Computational Modelling (ICM), University of Warsaw, under grant no. GB74-8. We thank Marek Tkacz for making available the laser drilling system at the Institute of Physical Chemistry of the Polish Academy of Sciences.

**Notes**: The authors declare no competing financial interest.

**Supporting Information:** (i) Comparison of calculated and experimental geometry of the $TiO_2$, $CaCl_2$, and HP-$PdF_2$ polymorphs of $ZnF_2$; (ii) Coefficients of the linear fits to the pressure dependence of the frequencies of the Raman-active vibrations; (iii) Comparison of the experimental Raman band position for the $CaCl_2$, HP-$PdF_2$, HP1-$AgF_2$, and $PbCl_2$ phases of $ZnF_2$; (iv) Evolution of the Raman spectrum of powdered $ZnF_2$ across the HP-$PdF_2$ to HP1-$AgF_2$ transition; (v) Comparison of the experimental Raman spectrum of $ZnF_2$ at 55 GPa with that simulated with LDA for the $PbCl_2$ polymorph; (vi) Comparison of the x-ray diffraction patter simulated for HP1-$AgF_2$ and $SrI_2$ structures optimized at 35 GPa; (vii) Comparison of the Raman spectrum simulated for HP1-$AgF_2$ and $SrI_2$ optimized at 35 GPa; (viii) Calculated crystal structure of the HP1-$AgF_2$ polymorph of $ZnF_2$ at 35 GPa (in VASP format).

# Supplementary Information

**Table S 1** Comparison of calculated and experimental geometry of the $TiO_2$, $CaCl_2$, and HP-$PdF_2$ polymorphs of $ZnF_2$ at selected pressures. Length of unit cell vectors ($a$, $b$, $c$) is given in Å, cell volume ($V$) in Å$^3$. Relative difference between experimental and theoretical values is given in parenthesis.

| Structure type | a | b | c | V | Ref. |
|---|---|---|---|---|---|
| $TiO_2$<br>p = 1 atm | 4.704 | | 3.134 | 69.33 | Exp. [53] |
| | 4.694 (–0.2 %) | | 3.122 (–0.2 %) | 68.81 (–0.8 %) | This work |
| | 4.712 (0.2 %) | | 3.179 (1.4 %) | 70.58 (1.8 %) | Calc. [59] |
| $CaCl_2$<br>p = 5.4 GPa | 4.612 | 4.751 | 3.065 | 67.16 | Exp. [10] |
| | 4.683 (1.5 %) | 4.502 (–5.2 %) | 3.097 | 65.30 (–2.8) | This work |
| HP-$PdF_2$<br>p = 15.3 GPa | | 4.881 | | 116.29 | Exp. [10] |
| | | 4.857 (–0.6 %) | | 114.58 (–1.5 %) | This work |

**Table S 2** Coefficients of the linear fits to the pressure dependence of the frequencies of the Raman-active vibrations.

| Phase | Mode symmetry | dν/dp (cm$^{-1}$/GPa) | ν$_{0GPa}$ (cm$^{-1}$) |
|---|---|---|---|
| $CaCl_2$ | $A_g$ | 3.6(1) | 357(1) |
| | $B_{1g} + B_{2g} + B_{3g}$ | 2.7(1) | 256(1) |
| | $A_g$ | 9.3(2) | 25(3) |
| HP-$PdF_2$ | $T_g$ | 4.02(6) | 441(1) |
| | $T_g$ | 2.50(6) | 344(1) |
| | $E_g$ | 3.81(3) | 243(1) |
| | $T_g$ | 2.53(8) | 214(2) |
| | $A_g$ | 1.93(3) | 174(1) |
| HP1-$AgF_2$ | $B_1$ | -0.08(8) | 421(3) |
| | $A_1$ | 0.51(8) | 341(3) |
| | $B_1$ | 1.5(2) | 262(7) |
| | $B_1$ | 1.9(2) | 212(6) |
| | $B_1$ | 1.5(1) | 180(4) |
| | $A_2$ | 1.33(8) | 151(3) |
| $PbCl_2$ | $B_{3g}$ | 1.5(4) | 323(19) |
| | $A_g$ | 0.8(2) | 293(9) |
| | $A_g$ | 1.9(1) | 273(7) |
| | $B_{2g}$ | 1.2(1) | 203(6) |
| | $A_g$ | 0.5(3) | 190(13) |
| | $A_g + B_{3g}$ | 0.65(4) | 126(2) |

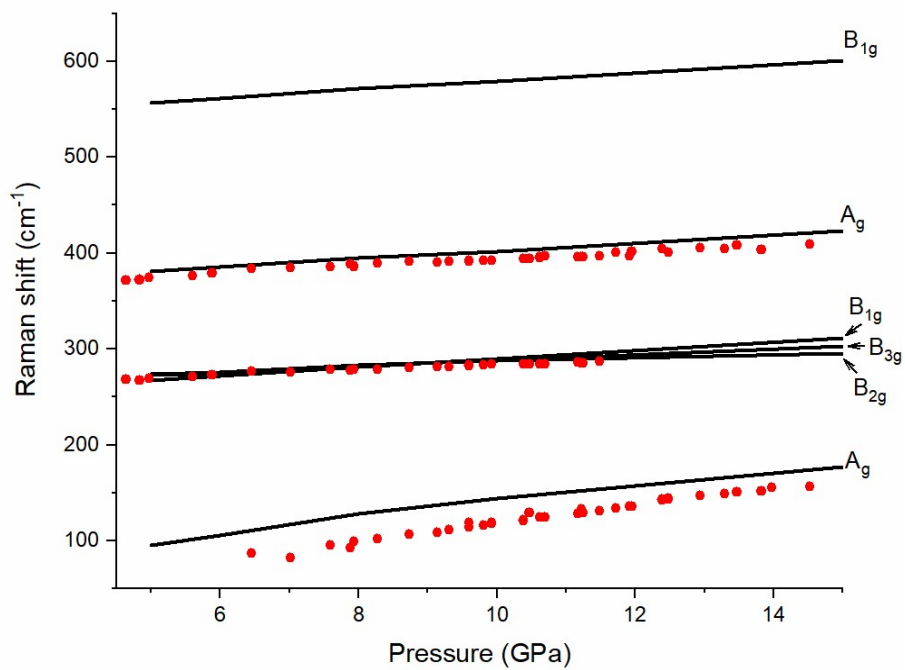

**Figure S 1** Comparison of the experimental Raman band position for the CaCl$_2$-type phase of ZnF$_2$ (red dots) with the frequencies of Raman-active Γ-point modes calculated for this polymorph using the SCAN functional (black lines). Labels denote the symmetry of each mode.

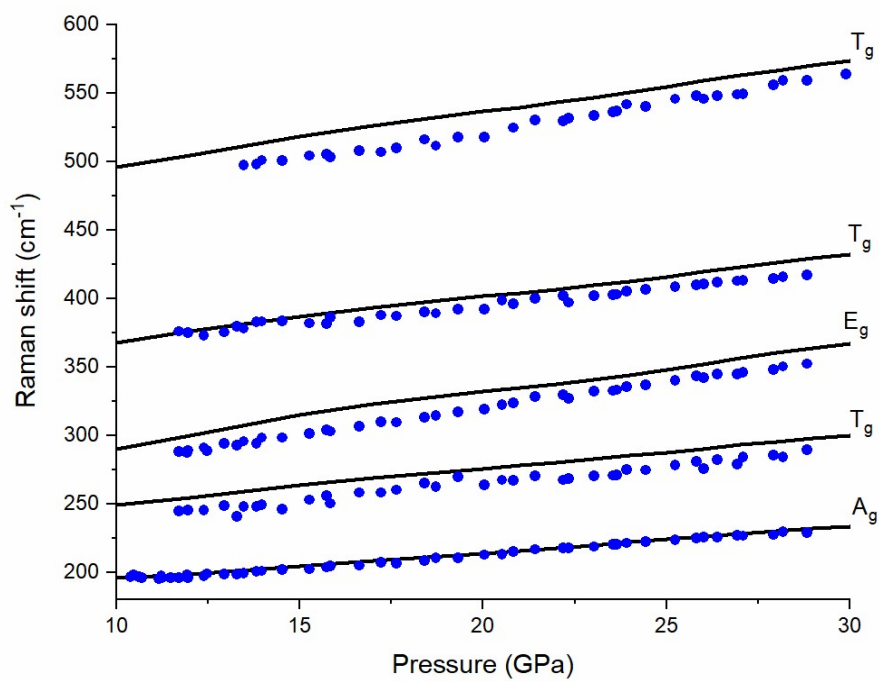

**Figure S 2** Comparison of the experimental Raman band position for the HP-PdF$_2$ phase of ZnF$_2$ (blue dots) with the frequencies of Raman-active Γ-point modes calculated for this polymorph using the SCAN functional (black lines). Labels denote the symmetry of each mode.

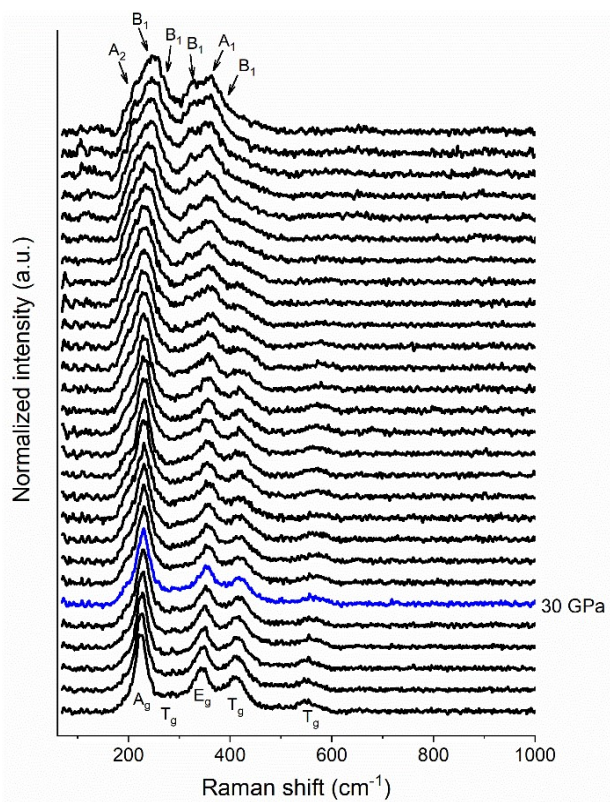

**Figure S 3** Evolution of the Raman spectrum of powdered $ZnF_2$ across the HP-$PdF_2$ to HP1-$AgF_2$ transition (26 to 43 GPa). Labels for the Raman bands of HP-$PdF_2$ (bottom) and HP1-$AgF_2$ (top) are given. The spectra are offset for clarity.

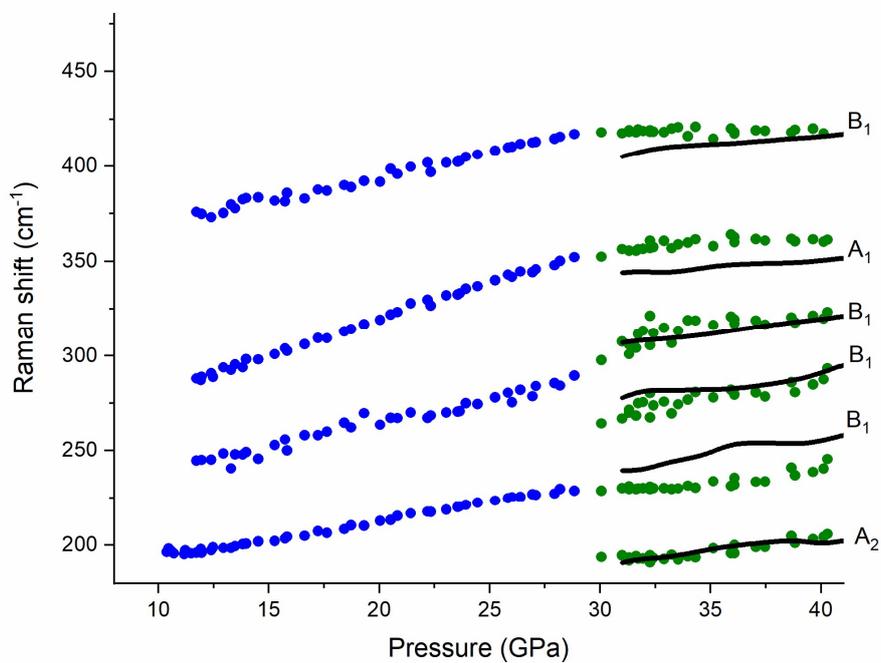

**Figure S 4** Comparison of the experimental Raman band position for the HP-$PdF_2$ and HP1-$AgF_2$ phases of $ZnF_2$ (blue and green dots) with the frequencies of selected Raman-active Γ-point modes calculated for HP1-$AgF_2$ polymorph using the SCAN functional (black lines). Labels denote the symmetry of each mode.

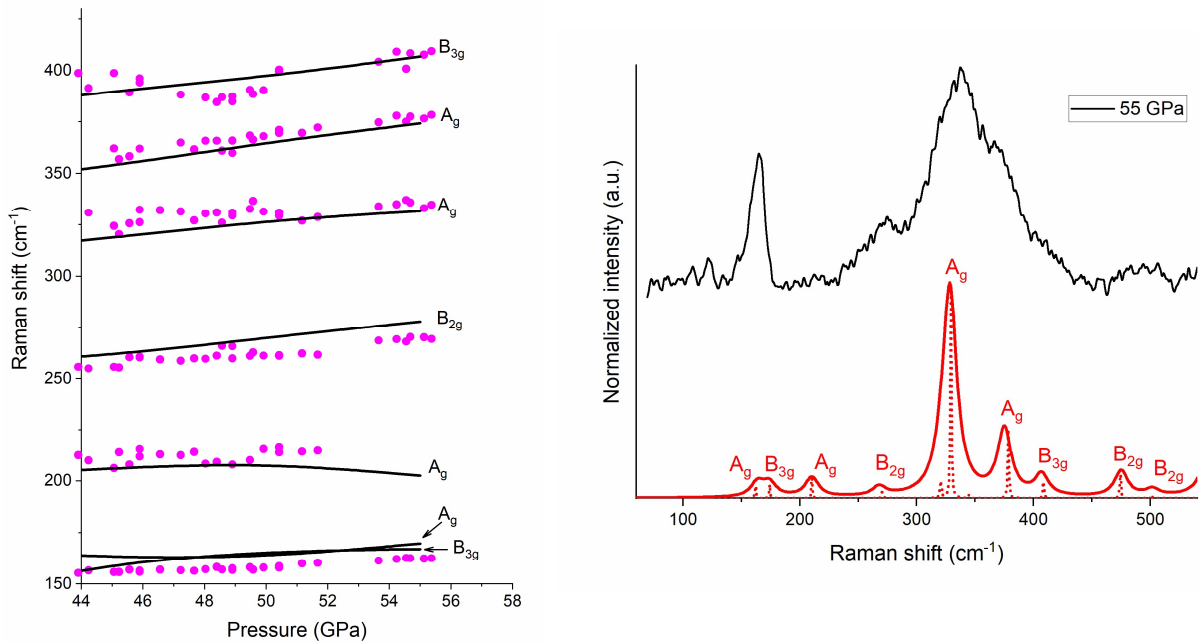

**Figure S 5 Left:** Comparison of the experimental Raman band position for the PbCl$_2$ phase of ZnF$_2$ (magenta dots) with the frequencies of the most intense Raman-active modes calculated for this polymorph using the SCAN functional (black lines). Labels denote the symmetry of each mode. **Right:** Comparison of the experimental Raman spectrum of ZnF$_2$ at 55 GPa with that simulated with LDA for the PbCl$_2$ polymorph. Red line denotes the spectrum simulated with a Lorentzian profile characterized by a full width at half maximum (FWHM) of 15 cm$^{-1}$. The dotted lines correspond to profiles with FHWM equal to 1 cm$^{-1}$.

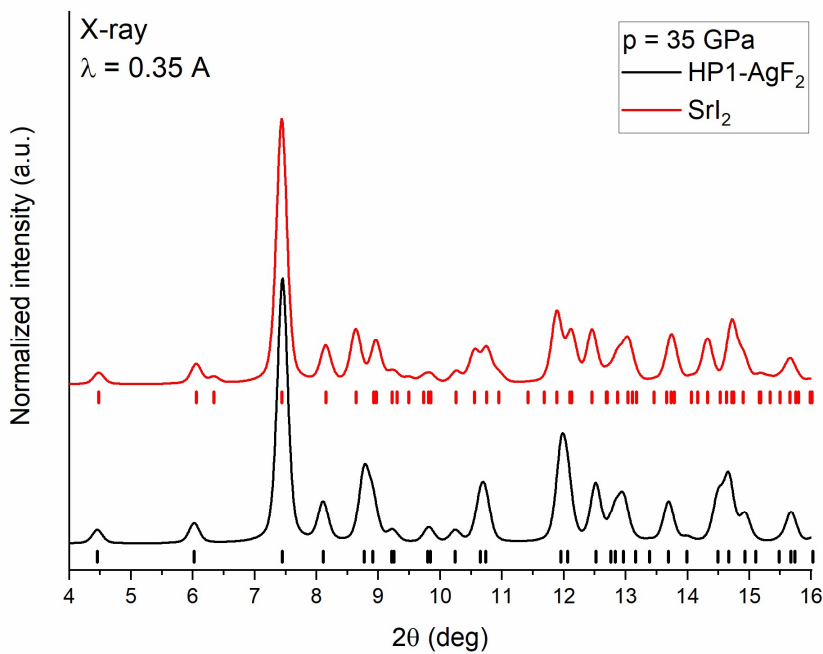

**Figure S 6** Comparison of the x-ray diffraction patter simulated for HP1-AgF$_2$ and SrI$_2$ structures optimized at 35 GPa. Vertical lines indicate position of Bragg peaks.

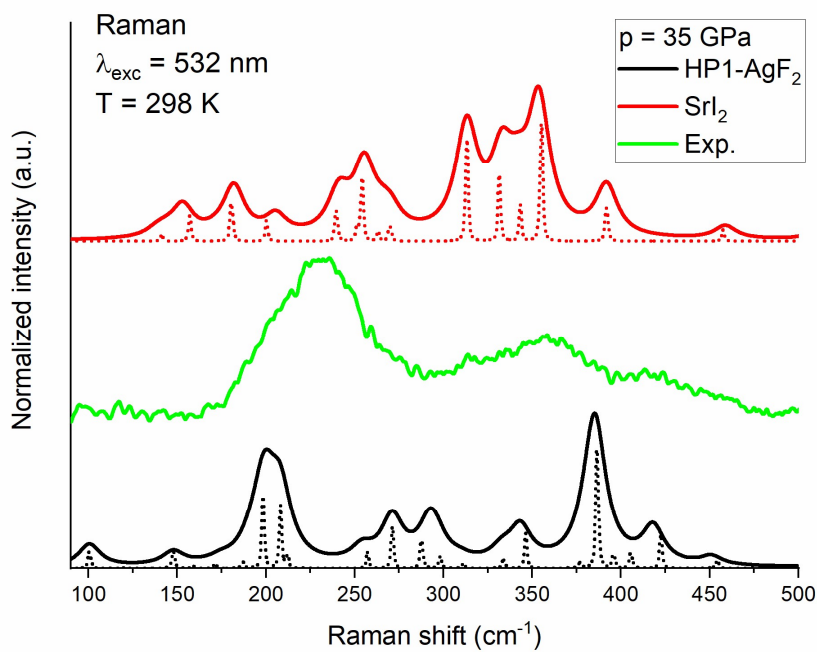

**Figure S 7** Comparison of the Raman spectrum simulated for HP1-AgF$_2$ and SrI$_2$ optimized at 35 GPa. Full lines indicate the spectrum simulated with a Lorentzian profile with a FWHM of 15 cm$^{-1}$. The dotted lines correspond to profiles with FHWM equal to 1 cm$^{-1}$.

**Calculated crystal structure of the HP1-AgF2 polymorph of ZnF2 at 35 GPa (in VASP format).**

```
HP1_AgF2 polymorph of ZnF2
  1.000000000000000
     4.6209453324633118    0.0000000000000000    0.0000000000000000
     0.0000000000000000    4.5857984157842093    0.0000000000000000
     0.0000000000000000    0.0000000000000000    4.8642234913989029
   Zn   F
    4    8
Direct
  0.9752991177817992  0.9913420485212134  0.9861618137091862
  0.9752991177817992  0.5286580094787874  0.4861618287091768
  0.4752991177817992  0.9913420485212134  0.5632781472908183
  0.4752991177817992  0.5286580094787874  0.0632781312908169
  0.3985438579427227  0.3697360649243133  0.4234472323536149
  0.7007269092754664  0.6849133546709573  0.7527392489136677
  0.7007269092754664  0.8350866893290424  0.2527392739136663
  0.3985438579427227  0.1502639040756837  0.9234472323536149
  0.8985438329427171  0.1502639040756837  0.6259927426463907
  0.2007269472754695  0.8350866893290424  0.2967007760863349
  0.2007269472754695  0.6849133546709573  0.7967007760863349
  0.8985438329427171  0.3697360649243133  0.1259927926463877
```